\documentclass[journal,twocolumn]{IEEEtran}

\usepackage{palatino}
\usepackage{mathtools}
\usepackage{url}
\usepackage{amsmath}
\usepackage{amssymb}
\usepackage{wrapfig}
\usepackage{framed}
\usepackage{mdframed}
\usepackage{verbatim}
 \usepackage{amsthm}
 \usepackage{cite}
\usepackage{subfigure}
\usepackage{palatino}
\usepackage{mathtools}
\usepackage{url}
\usepackage{amsmath}
\usepackage{amssymb}
\usepackage{wrapfig}
\usepackage{framed}
\usepackage{mdframed}
\usepackage{bbm}
\usepackage{verbatim}\usepackage{subfigure}
\usepackage[utf8]{inputenc}
\usepackage{amssymb}
\usepackage{amsthm}
\usepackage{dsfont}
\usepackage[normalem]{ulem}

\usepackage{bbm}
\usepackage{subfigure}
\usepackage{times}
\usepackage{multirow}
\usepackage[final]{graphicx}
\usepackage{upgreek}
\usepackage{latexsym,amssymb,mathrsfs}
\usepackage{comment}
\usepackage{url}

\usepackage{algorithmicx}
\usepackage[ruled]{algorithm}
\usepackage{algpseudocode}
\usepackage{algpascal}
\usepackage{algc}
\usepackage{xcolor}
\usepackage{cite}

 \usepackage{cite}

\title{SOK: Privacy Definitions and Classical Mechanisms in the Local Setting}
\author{\IEEEauthorblockN{Nan~ Wang, Likun~Qin, Tianshuo Qiu \\
}
\IEEEauthorblockA{Department of Electrical and Computer Engineering}\\
\IEEEauthorblockA{Shandong University, Jinan, China\\
}
}

\begin{document}
\maketitle
\begin{abstract}
This paper delves into the intricate landscape of privacy notions, specifically honed in on the local setting. Central to our discussion is the juxtaposition of point-wise protection and average-case protection, offering a comparative analysis that highlights the strengths and trade-offs inherent to each approach. Beyond this, we delineate between context-aware and context-free notions, examining the implications of both in diverse application scenarios. The study further differentiates between the interactive and non-interactive models, illuminating the complexities and nuances each model introduces. By systematically navigating these core themes, our goal is to provide a cohesive framework that aids researchers and practitioners in discerning the most suitable privacy notions for their specific requirements in the local setting.
\end{abstract}
\section{Introduction}
In the vast, interconnected expanse of our contemporary data ecosystem, the principles and mechanisms underpinning data privacy have never been more critical. As more sectors—from healthcare to commerce, from social networking to infrastructure—rely heavily on large-scale data processing, ensuring robust privacy, especially in the local setting, becomes a pivotal challenge. Within this challenge lies the multifaceted landscape of privacy notions, each tailored to cater to specific needs and scenarios.

Central to our exploration is the differentiation between point-wise protection and average-case protection\cite{duchi2013local, boukoros2019lack}. Point-wise protection, as the term suggests, focuses on the individual, ensuring that each data point, when taken in isolation, is afforded its unique protective shield against potential breaches or misuse. On the other hand, average-case protection operates on a broader scale, aiming to ensure that when data is looked at collectively or in aggregates, its privacy properties are maintained. The distinction between these two approaches is not merely theoretical; it has profound implications for how we design algorithms, systems, and policies.

Beyond this foundational dichotomy, the realm of privacy expands into the realms of context \cite{lopuhaa2019information}. Context-aware notions introduce an additional layer of complexity by ensuring that privacy definitions are crafted with a keen eye on the surrounding environment or setting of the data\cite{context}. It poses intriguing questions: How does the nature of a data-sharing platform, or the socio-cultural dynamics of a region, impact what 'privacy' truly means? Conversely, context-free notions strive for universality, aiming to provide privacy guarantees that remain robust, irrespective of the specific nuances of a given setting\cite{cao2019protecting}.

Last but not least, our discourse ventures into the interactive versus non-interactive models of privacy. The interactive model envisions a dynamic environment where the data and the querying mechanism engage in a continuous, two-way dialogue. This stands in stark contrast to the non-interactive model, where data is processed, analyzed, and subsequently shared or stored without further real-time interaction\cite{kairouz2016discrete}. This distinction, while subtle, is pivotal in determining the operational efficiency, user experience, and often, the very feasibility of certain privacy-preserving mechanisms.

Through the length and breadth of this study, our ambition extends beyond mere elucidation. We aim to bridge the gap between theory and practice, between abstract notions and tangible applications. By offering a comprehensive, nuanced, and layered understanding of privacy in the local setting, we hope to empower researchers, practitioners, policymakers, and even everyday users to navigate the intricate world of data privacy with confidence and clarity.

\begin{figure}
    \centering
    \includegraphics[width = 0.5\textwidth]{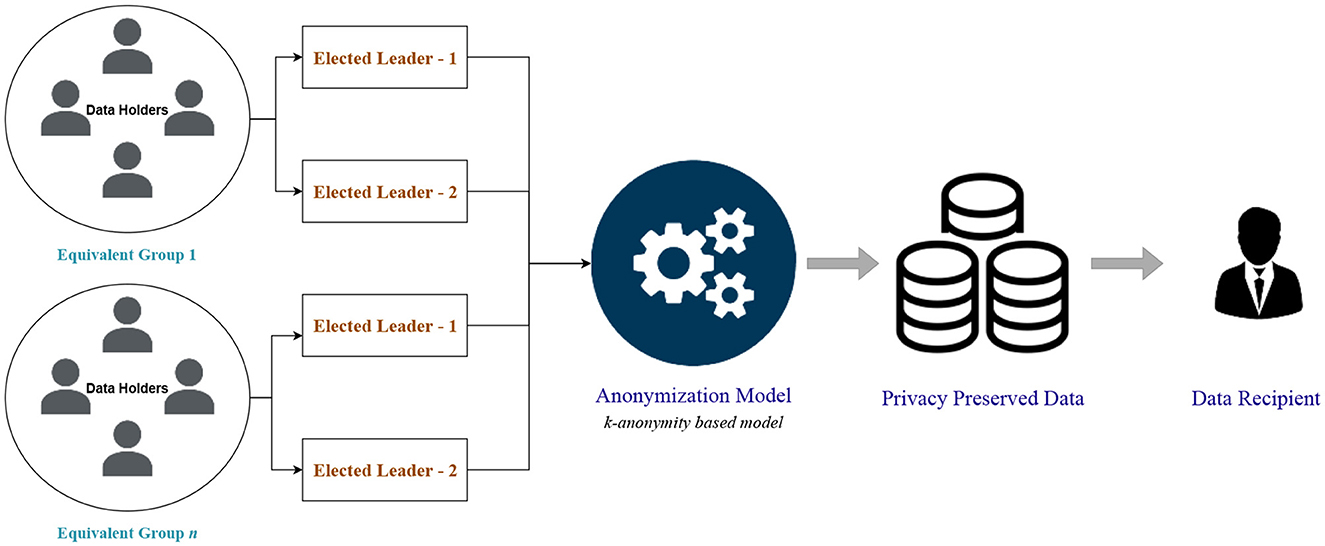}
    \caption{Model of local privacy notions}
    \label{fig:enter-label}
\end{figure}

\section{Delineating Point-wise and Average-case Protections}
The quest for privacy in data analysis begins with an understanding of the granularity at which protection is applied. This section dives deep into the nuances of two pivotal paradigms: point-wise protection and average-case protection.
\subsection{Point-wise Protection: Guarding Every Individual}
\textbf{Definition:} Point-wise protection refers to mechanisms that offer privacy guarantees for individual data points, ensuring that each unique entry is shielded from potential external breaches or internal misuse.

\textbf{Characteristics}: 1. Fine-grained Control: It allows for individualized protection settings, enabling users or systems to tweak privacy settings for specific data points.\cite{lopuhaa2020privacy}
2. Localized Errors: Since each data point is protected in isolation, any noise or error introduced for privacy does not propagate across the dataset.
3. Demands Higher Overheads: Given its granular nature, point-wise protection often demands higher computational and storage overheads.

In scenarios where the individual's data is extremely sensitive, like personal medical records or financial transactions, point-wise protection is pivotal

\subsection{Average-case Protection: Safeguarding the Collective} 
\textbf{Definition:} Average-case protection pertains to the privacy of data when viewed en masse. Instead of focusing on individual data points, it ensures that aggregates, statistics, or averages derived from the data are protected.

\textbf{Characteristics:}1. Broad Strokes: It is more about ensuring privacy for large groups rather than individuals, reducing the granularity of control.
2. Noise Distribution: Any noise introduced for privacy reasons is often distributed across the dataset, leading to more consistent and predictable errors.
3. Efficiency Gains: Due to its collective nature, average-case protection mechanisms can often be more efficient and scalable.

\subsection{Point-wise vs. Average-case: A Comparative Lens}

The interplay between point-wise and average-case protection forms the crux of many privacy-centric decisions in data analysis. Both methods, while aiming to offer privacy guarantees, follow inherently different paths, bringing distinct advantages and challenges. This section delves deeper into their comparative analysis.

Flexibility:
Point-wise protection stands out in its adaptability. By focusing on individual data points, it enables granular adjustments tailored to the unique sensitivities and characteristics of each data point. This granularity often translates to personalized privacy settings, ensuring that specific data points can be safeguarded based on their individual attributes. On the contrary, average-case protection casts a broader net. It looks at data in aggregates, ensuring that when a large set of data points is considered, privacy is not compromised. While this might not offer the individualized control that point-wise protection does, it brings a consistency that can be crucial for wide-scale data operations.

Overhead Costs:
The intricate nature of point-wise protection brings with it challenges. Given its focus on individual data points, computational resources are often stretched, leading to higher overheads both in terms of processing time and storage. This granularity can also mean that more intricate algorithms are required, especially when dealing with diverse datasets. Average-case protection, with its collective approach, often streamlines processes. By handling data en masse, it can introduce efficiencies, making it more scalable and often faster, especially when dealing with very large datasets.

Application Suitability:
The choice between point-wise and average-case often boils down to the end goal of the data operation. Point-wise protection is the go-to choice in situations where individual data points are of utmost importance. Think of applications like personal health records, where each record's confidentiality and integrity can have significant implications. Conversely, average-case finds its strength in scenarios where broader insights are the goal. When dealing with large-scale surveys, market research, or any application where discerning overarching trends is more vital than the specifics of individual data points, average-case protection comes into its own.

Data Utility vs. Privacy Trade-off:
An often-overlooked aspect is the trade-off between data utility and privacy. Point-wise, while offering heightened privacy for individual data points, might sometimes compromise on the overall utility of the data, especially when significant noise is introduced for privacy preservation. Average-case, by working on aggregates, often manages to retain a better balance, ensuring that while individual data points might be obfuscated, the overall insights drawn from the data remain intact.

Evolving Data Landscape:
As data operations evolve and datasets become more dynamic, the adaptability of the privacy mechanism becomes essential. Point-wise protection, with its granular approach, might be better suited to handle dynamic datasets where new data points are continuously added. In contrast, average-case, with its broader view, can be more resilient to minor changes, ensuring consistent privacy guarantees even as data evolves.

In essence, the debate between point-wise and average-case protection is not about which is superior, but rather about which is more appropriate for a given scenario. Recognizing the strengths and limitations of each is pivotal in making informed, effective, and efficient privacy-centric decisions.

\section{ Local Privacy Definitions: A Comparative Study}

\subsubsection{Local Differential Privacy (LDP)}
A mechanism \( M \) adheres to \(\epsilon\)-LDP if, for every input data point \( x \) and its possible adjacent data points \( x' \), the probability ratio of their outputs \( M(x) \) and \( M(x') \) remains bounded by \( e^{\epsilon} \).
\begin{equation}
\frac{P[M(x) = o]}{P[M(x') = o]} \leq e^{\epsilon} \quad \forall \, o\in Range(M)
\end{equation}

From the definition of LDP, we have the following observations: a) the likelihood ratio $\mathsf{LR}(y, x, x')$ of observing $y$ as an output for any two distinct inputs $x, x'$ from the mechanism $\mathcal{M}$ must be bounded; b) the definition does not assume any prior on the data; thus LDP provides prior-agnostic guarantees, and c) the likelihood probabilities, and thus the leakage of LDP only depends on the randomness used in the mechanism\cite{Cormode:2018:PSL:3183713.3197390}. 
 
To handle more general applications such as continuous-valued data release, a relaxed version of $\epsilon$-LDP is studied, which uses $\delta$ to measure the probability of the likelihood ratio is not bounded. Such relaxation is denoted as $(\epsilon,\delta)$-LDPc\cite{kairouz2014extremal, Dwork20061}.

Pros: Strong guarantees, known noise addition mechanisms. Cons: Can introduce significant noise, leading to reduced data utility.

\subsubsection{Local Randomized Response (LRR)}
In this mechanism, participants answer queries truthfully with a probability \( p \) and otherwise respond randomly with probability \( 1-p \).
\begin{equation}
\begin{aligned}
&P[\text{response}
=\text{true}]\\
= &p \cdot P[\text{truth}=\text{true}] + (1-p) \cdot P[\text{random}=\text{true}]
\end{aligned}
\end{equation}

Pros: Simple to implement, offers direct individual-level protection.
Cons: Not always optimal for more complex queries.

\subsubsection{Local k-Anonymity}
For every data point released, there exist at least \( k-1 \) other data points in the dataset that are indistinguishable from it. Given a dataset \( D \) and a released data point \( d \), the condition is:
\begin{equation}
|\{ d' \in D : \text{equivalence\_class}(d) = \text{equivalence\_class}(d') \}| \geq k
\end{equation}
Pros: Protects individual identity in the dataset.
Cons: Doesn’t necessarily protect against attribute disclosure.

\subsubsection{Local l-Diversity}
An extension of \( k \)-anonymity, ensuring that for each equivalence class, there are at least \( l \) “well-represented” values for the sensitive attribute. Formally, for every equivalence class \( E \) in the dataset:
\begin{equation}
|\{ \text{distinct sensitive\_values in } E \}| \geq l
\end{equation}

Pros: Protects against attribute disclosure.
Cons: More complex to achieve, potentially reducing data utility.

Now, we focus on context-aware privacy notions. Consider the setting where there is a prior distribution on the data, denoted by $\mathbf{P}$, and the prior is perfectly available when designing the mechanism. One such definition is mutual information privacy (MIP), which uses the mutual information between $Y$, $X$ to measure the average information leakage of $X$ contained in $Y$, MIP in the central setting takes $X$ as the input dataset and $Y$ as the output of the mechanism. Here, we twist it into the local setting, where $X$ denotes one user's local data in the following definition.

\subsubsection{Local Mutual Information Privacy (L-MIP)\cite{7498650}]}
A mechanism $\mathcal{M}$ satisfies $\epsilon$-L-MIP for some $\epsilon\in{\mathds{R}^+}$, if the mutual information between $X$ and $Y$ is bounded by $\epsilon$, i.e., $I(X;Y)\le{\epsilon}$.

Originally, MIP was proposed under the centralized setting where $X$ is the database or individual items, and $Y$ is a query output. Here we can adapt it to the local setting, where $X$ and $Y$ are each individual user's input and output. \cite{cuff2016differential}

\noindent L-MIP it is an average case leakage measure, i.e., the expected value of the log of the ratio of posterior $P(X=x|Y=y)$ to the prior $P(X=x)$. On the other hand, the average operation in L-MIP prohibits deriving closed-form perturbation parameters in mechanism design.

We next introduce Differential Identifiability. which provides context-aware pairwise protection overall possible values of $x$ and $y$.

\subsubsection{Local Information Privacy}

$[(\epsilon,\delta)$-Local Information Privacy]\cite{LIP2}
A mechanism $\mathcal{M}$  satisfies $(\epsilon,\delta)$-LIP for some $\epsilon\in{\mathds{R}^+}$ and $\delta\in[0,1]$, if $\forall{S_x\in{\mathcal{X}}}$, $S_y\in{\textit{Range}(\mathcal{M})}$:
\begin{equation}\label{cons0}
\begin{aligned}
    & P(Y\in\mathcal{S}_y) \geq e^{-\epsilon}P(Y\in{\mathcal{S}_y}|X\in\mathcal{S}_x)-\delta, \\
    &P(Y\in\mathcal{S}_y)\le{e^{\epsilon}P(Y\in{\mathcal{S}_y}|X\in\mathcal{S}_x)}+\delta.
\end{aligned}
\end{equation}

 \noindent The operational meaning of LIP is, the output $Y$ provides limited additional information about any possible input $X$, and the amount of the additional information is measured by the privacy budget $\epsilon$ and failure probability $\delta$. Note that,  when $\epsilon$ is small, the posterior probability of $X$ given $Y$ is close to the prior of $X$\cite{jiang2019local}.



\subsubsection{$\epsilon$-Differential Identifiability (DI)} \cite{Lee:2012:DI:2339530.2339695,7498650}
A mechanism $\mathcal{M}$ satisfies  $\epsilon$-DI for some $\epsilon\in{\mathds{R}^+}$, if $\forall{S_x, S_x' \in{\mathcal{X}}}$ and $\forall{S_y\in{\textit{Range}(\mathcal{M})}}$ if \begin{equation*}
    \frac{P(X\in\mathcal{S}_x|Y\in\mathcal{S}_y)}{P(X\in\mathcal{S}_x'|Y\in\mathcal{S}_y)}\le{e^{\epsilon}}.
\end{equation*}

\subsubsection{$\epsilon$-Local Pufferfish Privacy \cite{kifer2012rigorous}}
Given set of potential secrets $S$, a set of discriminative pairs $S_{pairs}$, a set of data evolution scenarios $\mathsf{\theta}_{\mathcal{X},\mathcal{S}} $, and a privacy parameter $\epsilon\in{\mathds{R}^+}$, an (potentially randomized) algorithm $\mathcal{M}$ satisfies $\epsilon$-Pufferfish ($S$, $S_{pairs}$, $\mathsf{\theta}_{\mathcal{X},\mathcal{S}} $) privacy if
for all possible outputs $y \in \text{range}(\mathcal{M})$,
 for all pairs $(s_i,s_j)\in S_{pairs}$ of potential secrets,
for all distributions $\theta_{\mathcal{X},\mathcal{S}}\in \mathsf{\theta}_{\mathcal{X},\mathcal{S}} $ for which $P(s_i| \theta_{\mathcal{X},\mathcal{S}})\neq{0}$ and $P(s_j|\theta_{\mathcal{X},\mathcal{S}})\neq{0}$, the following holds:
\begin{equation*}
\frac{P(\mathcal{M}(X)=y|\theta_{\mathcal{X},\mathcal{S}},s_i)}{P(\mathcal{M}(X)=y|\theta_{\mathcal{X},\mathcal{S}},s_j)}\le{e^{\epsilon}}.
\end{equation*}
\noindent The first observation is that the pufferfish privacy is not necessarily context-aware, as it defines a bounded region for all the possible prior distributions, when the region is large enough to include all possible priors, it is equivalent to a context-free setting. Also note that, the input $X$ is not necessarily sensitive itself, but may be correlated to some hidden properties defined in the secret set, the correlations are described through $\mathsf{\theta}_{\mathcal{X},\mathcal{S}} $, i.e., $\mathsf{\theta}_{\mathcal{X},\mathcal{S}} $ is not only the set of all possible prior distribution, but also contains the data dependence\cite{song2017pufferfish}.

The next privacy notion is Maximal Information Leakage (MIL). Suppose an adversary is interested in some sensitive data $U$ with domain $\mathcal{U}$ which is correlated to $X$, then observing $Y$, MIL measures the information gain at the adversary from the aspect of correct guessing:

\subsubsection{$\epsilon$-Maximal Information Leakage (MIL)\cite{DBLP:journals/corr/abs-1807-07878} } The maximal information leakage for a mechanism $\mathcal{M}$ is
\begin{align}\label{MIL_C}
     \mathsf{L}_{\textsf{MIL}}(X;{Y})=\sup_{U-X-Y-\hat{U}}\ln \frac{P({U=\hat{U}})}{\max_{u\in\mathcal{U}}P_U(u)},
\end{align}
{where $U$ is a (possibly randomized) function of $X$, $\hat{U}$ denotes a guess from the adversary.} For any joint distribution $P_{X,Y}$ on finite alphabets $\mathcal{X}$ and $\mathcal{Y}$, \eqref{MIL_C} can be rewritten as

\begin{align}
 \mathsf{L}_{\textsf{MIL}}(X;{Y})=\ln \sum_{y\in{\mathcal{Y}}}\max_{x\in{\mathcal{X}}}P_{Y|X}(y|x) \nonumber
\end{align}
and $\mathcal{M}$ satisfies $\epsilon$-Maximal Information Leakage Privacy if $\mathsf{L}_{\textsf{MIL}}(X;{Y}) \leq \epsilon $. 
if for some $\epsilon\in{\mathds{R}^+}$, $\forall{x\in{\mathcal{X}}}$, $\forall{y\in\mathcal{Y}}$.

\noindent MIL captures the average likelihood over all possible $y\in{\mathcal{Y}}$ given the corresponding value of $x$(s) that provide the maximal likelihood probability. However, MIL does not provide pairwise protection over all possible values of $x$ and $y$.

Thus, for localized pufferfish privacy, we have, $\forall{s_i,s_j\in{S_{pair}}}$, ${P}_X\in \mathsf{\theta}_{\mathcal{X}} $, and  $y \in \text{range}(\mathcal{M})$:
\begin{equation}\label{eq2}
   e^{-\epsilon}\le\frac{P(\mathcal{M}(s_i)=y|{P}_X)}{P(\mathcal{M}(s_j)=y|{P}_X)}\le{e^{\epsilon}}.
\end{equation}

One drawback of pufferfish privacy is the lack of mechanisms, even though in \cite{DBLP:journals/corr/WangSC16}, a Wasserstein Mechanism is proposed, however, it is computationally impractical, and the approximated mechanism trades in some utility. 

\subsubsection{Geo-indistinguishability}
Geo-indistinguishability\cite{Geo} is a privacy measure that ensures that the location of an individual remains indistinguishable within a certain geographical radius. Formally, a mechanism \(M\) ensures \(\epsilon\)-geo-indistinguishability if for any two locations \(x, x'\) and any possible output \(o\), the following holds:

\begin{equation}
\frac{P[M(x) = o]}{P[M(x') = o]} \leq e^{\epsilon \cdot d(x, x')}
\end{equation}

where \(d(x, x')\) is the Euclidean distance between the two locations.

 \noindent The operational meaning of LIP is, the output $Y$ provides limited additional information about any possible input $X$, and the amount of the additional information is measured by the privacy budget $\epsilon$ and failure probability $\delta$. Note that,  when $\epsilon$ is small, the posterior probability of $X$ given $Y$ is close to the prior of $X$. 


\section{Classical Mechanisms for Each Privacy Notion}

In the evolving domain of data protection, various mechanisms have emerged to guarantee privacy to users. Each of these mechanisms aligns with specific privacy notions, ensuring optimal utility while preventing unintended information leakage. This section offers a detailed examination of some of the significant privacy mechanisms available:

Randomized Response (RR) Mechanism:

This mechanism involves introducing randomness in responses to mask the actual data values. Examples including RAPPOR from Google \cite{Rappor}.  Suitable for Local Differential Privacy (LDP)\cite{joseph2018local, LDP1}, Local Information Privacy (LIP), and other privacy paradigms that emphasize pointwise privacy protection. By using randomization, this method ensures that the output is not directly associated with the input, thus providing a certain degree of privacy.

Random Sampling Mechanism:
This mechanism is based on the probabilistic release of actual data values. Data is released in its true form with a probability $p$ and is sampled from the underlying data distribution with probability $1-p$. Essentially, this mechanism either reveals the actual data or provides a sample that resembles it. With a given prior distribution, it can be translated into the Randomized Response (RR) mechanism\cite{Locationprivacy}.

Convex Optimization Mechanism:
Designed for Maximal Information Privacy (MIP)\cite{9715086}, this mechanism is grounded in the principle that the privacy definition is convex. By leveraging convex optimization techniques, it aims to find the best protection strategy while maintaining utility and ensuring that the data's privacy is not compromised.

Wasserstein Mechanism:
A unique approach primarily used for Pufferfish privacy, the Wasserstein mechanism determines the distance between two data distributions. By leveraging the Wasserstein distance, it ensures that any two given distributions are indistinguishable to a certain extent, thus providing a robust privacy guarantee.

Noise Adding Mechanism:
One of the fundamental methods of ensuring data privacy is by adding noise to the data\cite{bound}. Common noise distributions include Gaussian and Laplacian. However, there are challenges in the local model, especially in measuring sensitivity. The primary drawback is the often undesirable trade-off between utility and privacy. When too much noise is added, the utility of the data diminishes, and vice versa.

Utility Optimized Mechanism\cite{qin2016heavy,murakami2019utility}:
Recognizing the essential balance between utility and privacy, some mechanisms have been developed to optimize utility without significantly compromising privacy. \cite{xu2018distilling,han2015minimax,bassily2017practical} Examples include Optimized Local Differential Privacy (OLDP), Optimized Unary Encoding Local Differential Privacy (OUE-LDP)\cite{Tianhao}, and Optimized Unary Encoding Local Information Privacy (OUE-LIP)\cite{LIP1}. These mechanisms employ optimization techniques to find the right balance, ensuring that data remains useful while maintaining the desired privacy level.

\section{Conclusion and Future Works:}
This paper delved deep into the intricate world of privacy definitions, examining their nuances and presenting mechanisms tailored to each. From the classic Randomized Response to the more contemporary Utility Optimized Mechanisms, our exploration underscores a profound commitment in the academic and tech community towards securing individual data.

Our study has revealed two primary insights. Firstly, there is no one-size-fits-all approach in the realm of data protection. Depending on the unique requirements and constraints of each application, different mechanisms and privacy definitions may be more apt. Secondly, the dynamic between data utility and privacy stands as a central challenge in this field. As mechanisms become more sophisticated, ensuring that they don't unduly compromise the utility of data becomes increasingly paramount.

Future research in this domain should focus on the interplay of these mechanisms in real-world scenarios, quantifying their efficiency in diverse datasets and applications. Furthermore, as threats to privacy evolve, so must our mechanisms, making this an ever-evolving field of study.

\bibliographystyle{IEEEtran}
\bibliography{ref}

\end{document}